\documentstyle[12pt,epsf]{article}

\catcode`\@=11
\long\def\@makefntext#1{ 
\protect\noindent \hbox to 3.2pt {\hskip-.9pt
$^{{\ninerm\@thefnmark}}$\hfil}#1\hfill} 

\def\thefootnote{\fnsymbol{footnote}}
 \def\@makefnmark{\hbox to 0pt{$^{\@thefnmark}$\hss}}  

\def\ps@myheadings{\let\@mkboth\@gobbletwo
\def\@oddhead{\hbox{} 
\rightmark\hfil\ninerm\thepage}
\def\@oddfoot{}\def\@evenhead{\ninerm\thepage\hfil 
\leftmark\hbox{}}\def\@evenfoot{}
\def\sectionmark##1{}\def\subsectionmark##1{}}

\begin{document}

\newcommand{\symbolfootnote}{\renewcommand{\thefootnote}
	{\fnsymbol{footnote}}}
\renewcommand{\thefootnote}{\fnsymbol{footnote}}
\newcommand{\alphfootnote}
	{\setcounter{footnote}{0}
	 \renewcommand{\thefootnote}{\sevenrm\alph{footnote}}}

\newcounter{sectionc}\newcounter{subsectionc}\newcounter{subsubsectionc}
\renewcommand{\section}[1] {\vspace{0.6cm}\addtocounter{sectionc}{1}
\setcounter{subsectionc}{0}\setcounter{subsubsectionc}{0}\noindent
	{\bf\thesectionc. #1}\par\vspace{0.4cm}}
\renewcommand{\subsection}[1] {\vspace{0.6cm}\addtocounter{subsectionc}{1}
	\setcounter{subsubsectionc}{0}\noindent
	{\it\thesectionc.\thesubsectionc. #1}\par\vspace{0.4cm}}
\renewcommand{\subsubsection}[1]
{\vspace{0.6cm}\addtocounter{subsubsectionc}{1}
	\noindent {\rm\thesectionc.\thesubsectionc.\thesubsubsectionc.
	#1}\par\vspace{0.4cm}}
\newcommand{\nonumsection}[1] {\vspace{0.6cm}\noindent{\bf #1}
	\par\vspace{0.4cm}}

\newcounter{appendixc}
\newcounter{subappendixc}[appendixc]
\newcounter{subsubappendixc}[subappendixc]
\renewcommand{\thesubappendixc}{\Alph{appendixc}.\arabic{subappendixc}}
\renewcommand{\thesubsubappendixc}
	{\Alph{appendixc}.\arabic{subappendixc}.\arabic{subsubappendixc}}

\renewcommand{\appendix}[1] {\vspace{0.6cm}
        \refstepcounter{appendixc}
        \setcounter{figure}{0}
        \setcounter{table}{0}
        \setcounter{equation}{0}
        \renewcommand{\thefigure}{\Alph{appendixc}.\arabic{figure}}
        \renewcommand{\thetable}{\Alph{appendixc}.\arabic{table}}
        \renewcommand{\theappendixc}{\Alph{appendixc}}
        \renewcommand{\theequation}{\Alph{appendixc}.\arabic{equation}}
        \noindent{\bf Appendix \theappendixc #1}\par\vspace{0.4cm}}
\newcommand{\subappendix}[1] {\vspace{0.6cm}
        \refstepcounter{subappendixc}
        \noindent{\bf Appendix \thesubappendixc. #1}\par\vspace{0.4cm}}
\newcommand{\subsubappendix}[1] {\vspace{0.6cm}
        \refstepcounter{subsubappendixc}
        \noindent{\it Appendix \thesubsubappendixc. #1}
	\par\vspace{0.4cm}}

\def\abstracts#1{{
	\centering{\begin{minipage}{30pc}\tenrm\baselineskip=12pt\noindent
	\centerline{\tenrm ABSTRACT}\vspace{0.3cm}
	\parindent=0pt #1
	\end{minipage} }\par}}

\def\adabstracts#1{{
	\centering{\begin{minipage}{30pc}\tenrm\baselineskip=12pt\noindent
	\parindent=0pt #1
	\end{minipage} }\par}}

\newcommand{\bibit}{\it}
\newcommand{\bibbf}{\bf}
\renewenvironment{thebibliography}[1]
	{\begin{list}{\arabic{enumi}.}
	{\usecounter{enumi}\setlength{\parsep}{0pt}
\setlength{\leftmargin 1.25cm}{\rightmargin 0pt}
	 \setlength{\itemsep}{0pt} \settowidth
	{\labelwidth}{#1.}\sloppy}}{\end{list}}

\topsep=0in\parsep=0in\itemsep=0in
\parindent=1.5pc

\newcounter{itemlistc}
\newcounter{romanlistc}
\newcounter{alphlistc}
\newcounter{arabiclistc}
\newenvironment{itemlist}
    	{\setcounter{itemlistc}{0}
	 \begin{list}{$\bullet$}
	{\usecounter{itemlistc}
	 \setlength{\parsep}{0pt}
	 \setlength{\itemsep}{0pt}}}{\end{list}}

\newenvironment{romanlist}
	{\setcounter{romanlistc}{0}
	 \begin{list}{$($\roman{romanlistc}$)$}
	{\usecounter{romanlistc}
	 \setlength{\parsep}{0pt}
	 \setlength{\itemsep}{0pt}}}{\end{list}}

\newenvironment{alphlist}
	{\setcounter{alphlistc}{0}
	 \begin{list}{$($\alph{alphlistc}$)$}
	{\usecounter{alphlistc}
	 \setlength{\parsep}{0pt}
	 \setlength{\itemsep}{0pt}}}{\end{list}}

\newenvironment{arabiclist}
	{\setcounter{arabiclistc}{0}
	 \begin{list}{\arabic{arabiclistc}}
	{\usecounter{arabiclistc}
	 \setlength{\parsep}{0pt}
	 \setlength{\itemsep}{0pt}}}{\end{list}}

\newcommand{\fcaption}[1]{
        \refstepcounter{figure}
        \setbox\@tempboxa = \hbox{\tenrm Fig.~\thefigure. #1}
        \ifdim \wd\@tempboxa > 6in
           {\begin{center}
        \parbox{6in}{\tenrm\baselineskip=12pt Fig.~\thefigure. #1 }
            \end{center}}
        \else
             {\begin{center}
             {\tenrm Fig.~\thefigure. #1}
              \end{center}}
        \fi}

\newcommand{\tcaption}[1]{
        \refstepcounter{table}
        \setbox\@tempboxa = \hbox{\tenrm Table~\thetable. #1}
        \ifdim \wd\@tempboxa > 6in
           {\begin{center}
        \parbox{6in}{\tenrm\baselineskip=12pt Table~\thetable. #1 }
            \end{center}}
        \else
             {\begin{center}
             {\tenrm Table~\thetable. #1}
              \end{center}}
        \fi}

\def\@citex[#1]#2{\if@filesw\immediate\write\@auxout
	{\string\citation{#2}}\fi
\def\@citea{}\@cite{\@for\@citeb:=#2\do
	{\@citea\def\@citea{,}\@ifundefined
	{b@\@citeb}{{\bf ?}\@warning
	{Citation `\@citeb' on page \thepage \space undefined}}
	{\csname b@\@citeb\endcsname}}}{#1}}

\newif\if@cghi
\def\cite{\@cghitrue\@ifnextchar [{\@tempswatrue
	\@citex}{\@tempswafalse\@citex[]}}
\def\citelow{\@cghifalse\@ifnextchar [{\@tempswatrue
	\@citex}{\@tempswafalse\@citex[]}}
\def\@cite#1#2{{$\null^{#1}$\if@tempswa\typeout
	{IJCGA warning: optional citation argument
	ignored: `#2'} \fi}}
\newcommand{\citeup}{\cite}

\def\fnm#1{$^{\mbox{\scriptsize #1}}$}
\def\fnt#1#2{\footnotetext{\kern-.3em
	{$^{\mbox{\sevenrm #1}}$}{#2}}}

\font\twelvebf=cmbx10 scaled\magstep 1
\font\twelverm=cmr10 scaled\magstep 1
\font\twelveit=cmti10 scaled\magstep 1
\font\elevenbfit=cmbxti10 scaled\magstephalf
\font\elevenbf=cmbx10 scaled\magstephalf
\font\elevenrm=cmr10 scaled\magstephalf
\font\elevenit=cmti10 scaled\magstephalf
\font\bfit=cmbxti10
\font\tenbf=cmbx10
\font\tenrm=cmr10
\font\tenit=cmti10
\font\ninebf=cmbx9
\font\ninerm=cmr9
\font\nineit=cmti9
\font\eightbf=cmbx8
\font\eightrm=cmr8
\font\eightit=cmti8
\newlength{\extraspace}
\setlength{\extraspace}{2mm}
\newlength{\extraspaces}
\setlength{\extraspaces}{2.0mm}
\newcommand{\be}{\begin{equation}}
\addtolength{\abovedisplayskip}{\extraspaces}
\addtolength{\belowdisplayskip}{\extraspaces}
\addtolength{\abovedisplayshortskip}{\extraspace}
\addtolength{\belowdisplayshortskip}{\extraspace}
\newcommand{\ee}{\end{equation}}
\newcommand{\bear}{\begin{eqnarray}}
\addtolength{\abovedisplayskip}{\extraspaces}
\addtolength{\belowdisplayskip}{\extraspaces}
\addtolength{\abovedisplayshortskip}{\extraspace}
\addtolength{\belowdisplayshortskip}{\extraspace}
\newcommand{\eear}{\end{eqnarray}}
\newcommand{\ba}{\begin{array}}
\newcommand{\ea}{\end{array}}
\newcommand{\vev}{\mbox{$\langle \rho \rangle$}}
\newcommand{\trace}{\mbox{\rm Tr}}
\newcommand{\ie}{{\it i.e.\ }}
\newcommand{\cf}{{\it c.f. }}
\newcommand{\eg}{{\it e.g.\ }}
\newcommand{\rtoo}{\mbox{$\sqrt{2}$}}
\newcommand{\pauli}{\mbox{$\vec{\tau}$}}
\newcommand{\ww}{\mbox{$W_{L}W_{L}$}}
\newcommand{\zz}{\mbox{$Z_{L}Z_{L}$}}
\newcommand{\vw}{\mbox{$\vec{w}$}}
\newcommand{\vpse}{\mbox{$\vec{p}$}}
\newcommand{\vpi}{\mbox{$\vec{\pi}$}}
\newcommand{\pdmu}{\mbox{$\partial_{\mu}$}}
\newcommand{\PDmu}{\mbox{$\partial^{\mu}$}}
\newcommand{\su}{\mbox{$SU(2)\times U(1)$}}
\newcommand{\stwo}{\mbox{$SU(2)\times SU(2)$}}
\newcommand{\lr}{\mbox{$SU(2)_{L}\times SU(2)_{R}$}}
\newcommand{\ew}{\mbox{$SU(2)_{W}\times U(1)_{Y}$}}
\newcommand{\hc}{{\rm h.c.}}
\newcommand{\gev}{\mbox{\rm GeV}}
\newcommand{\wz}{\mbox{$w^{+} w^{-} \rightarrow z z$}}
\newcommand{\lpr}{\mbox{$\lambda^{\prime}$}}
\newcommand{\yu}{\mbox{$\Sigma$}}
\newcommand{\xee}{\mbox{$\xi$}}
\newcommand{\xpr}{\mbox{$\xi^{\prime}$}}
\newcommand{\xdpr}{\mbox{$\xi^{\prime\prime}$}}
\newcommand{\ctri}{\mbox{$\lambda_{3}$}}
\newcommand{\ctes}{\mbox{$\lambda_{4}$}}
\newcommand{\cA}{{\cal A}} \newcommand{\cN}{{\cal N}}
\newcommand{\cB}{{\cal B}} \newcommand{\cO}{{\cal O}}
\newcommand{\cC}{{\cal C}} \newcommand{\cP}{{\cal P}}
\newcommand{\cD}{{\cal D}} \newcommand{\cQ}{{\cal Q}}
\newcommand{\cE}{{\cal E}} \newcommand{\cR}{{\cal R}}
\newcommand{\cF}{{\cal F}} \newcommand{\cS}{{\cal S}}
\newcommand{\cG}{{\cal G}} \newcommand{\cT}{{\cal T}}
\newcommand{\cH}{{\cal H}} \newcommand{\cU}{{\cal U}}
\newcommand{\cI}{{\cal I}} \newcommand{\cV}{{\cal V}}
\newcommand{\cJ}{{\cal J}} \newcommand{\cW}{{\cal W}}
\newcommand{\cK}{{\cal K}} \newcommand{\cX}{{\cal X}}
\newcommand{\cL}{{\cal L}} \newcommand{\cY}{{\cal Y}}
\newcommand{\cM}{{\cal M}} \newcommand{\cZ}{{\cal Z}}
\newcommand{\np}{Nucl.\ Phys.\ {\bf B}}
\newcommand{\pr}{Phys.\ Rev.\ }
\newcommand{\prd}{Phys.\ Rev.\ {\bf D}}
\newcommand{\prp}{Phys.\ Rep.\ }
\newcommand{\prl}{Phys.\ Rev.\ Lett.\ }
\newcommand{\pl}{Phys.\ Lett.\ {\bf B}}
\newcommand{\jsp}{J.\ Stat.\ Phys.\ }
\newcommand{\cmp}{Comm.\ Math.\ Phys.\ }
\newcommand{\zp}{Z. Phys.\ {\bf C}}
\newcommand{\ptp}{Prog.\ Theor.\ Phys.\ }
\newcommand{\ap}{Ann.\ Phys.\ }


\centerline{\tenbf WHEN IS A HIGGS {\underline{THE}} HIGGS?}
\baselineskip=22pt
\centerline{\tenbf OR }
\baselineskip=22pt
\centerline{\tenbf THE PHENOMENOLOGY OF A NON-STANDARD HIGGS BOSON}
\vspace{0.8cm}
\centerline{\tenrm R. SEKHAR CHIVUKULA}
\vspace{0.3cm}
\centerline{\tenrm and}
\vspace{0.3cm}
\centerline{\tenrm VASSILIS KOULOVASSILOPOULOS}
\vspace{0.3cm}
\baselineskip=13pt
\centerline{\tenit Physics Department, Boston University}
\baselineskip=12pt
\centerline{\tenit 590 Commonwealth Ave.}
\baselineskip=12pt
\centerline{\tenit Boston, MA 02215}
\vspace{0.9cm}
\abstracts{
The one-Higgs-doublet standard model is necessarily incomplete because of
  the triviality of the scalar symmetry-breaking sector.  If the Higgs mass is
  approximately 600 GeV or higher, there must be additional dynamics at a
  scale $\Lambda$ which is less than a few TeV. In this case the properties of
  the Higgs resonance can differ substantially from those predicted by the
  standard model.  In this talk we construct an effective Lagrangian
  description of a theory with a non-standard Higgs boson and analyze the
  features of a theory with such a resonance coupled to the Goldstone Bosons
  of the breaking of $\su$.
}
\vspace{0.3cm}
\adabstracts{\it
Talk presented by R.~S.~Chivukula at the conference on ``Continuous
Advances in QCD'', Minneapolis, Feb. 18-20, 1994. }
\vspace{0.3cm}
\hfil{\it BUHEP-94-9}

\hfil{\tt hep-ph/9405275}

\vspace{0.8cm}
\rm\baselineskip=14pt
\section{Apologia}

Let me begin by stating the reason that I believe a talk of this sort
has a place in a conference on QCD.  QCD, in addition to its intrinsic
interest and relevance, is a prototype of a theory with a dynamically
broken chiral symmetry. It is, therefore, a prototype of a theory of
dynamical electroweak symmetry breaking. The simplest example of a
theory with dynamical electroweak symmetry, Technicolor$^1$, is a
vector-like $SU(N)$ gauge theory, just like QCD, with a dynamical
scale of approximately 1 TeV instead of 1 GeV. Unfortunately, this
type of QCD-like theory is disfavored for a number of phenomenological
reasons$^2$. It is important, therefore, to investigate and understand
other members of the class of theories which allow for dynamical
electroweak symmetry breaking. This investigation will clearly employ
some of the methods ({\it e.g.} Chiral Perturbation Theory) which have
been successfully applied to QCD. Furthermore, progress in
understanding the dynamics of chiral symmetry breaking in QCD may
allow us to construct new, phenomenologically acceptable, theories of
dynamical electroweak symmetry breaking. I hope that the QCD experts
gathered here will keep this last point in mind.

The standard one-doublet Higgs model of the weak interactions
is in spectacular agreement with experimental results. However:
\begin{itemize}
\item There is no direct experimental evidence for the existence
of a Higgs Boson.
\item (Non-supersymmetric) Theories with fundamental scalars
are unnatural.
\end{itemize}
Furthermore, the symmetry-breaking sector of the standard
model is trivial$^3$. That is, the theory can only be understood
as a low-energy effective theory for some, more fundamental,
high-energy theory.

If the mass of the Higgs Boson is greater than about 600 -- 700
GeV, the scale $\Lambda$ of the new dynamics cannot be greater$^4$
than of order one or a few TeV. Turning this last argument
around: if the electroweak symmetry breaking sector involves
a heavy (iso-)scalar resonance that couples to the electroweak
gauge Bosons, then (since the scale of the new dynamics
is relatively low) this particle will likely have properties
rather different from those of the SM Higgs Boson. We
call such a resonance a ``non-standard'' Higgs Boson$^5$.

As an existence proof, note that there are (at least) two classes of
dynamical electroweak symmetry breaking models known where a
``non-standard'' Higgs may be present:
\begin{itemize}
\item Georgi-Kaplan Composite Higgs models$^6$: in these
models, all four members of a Higgs doublet are Goldstone Bosons
arising from chiral symmetry breaking due to a strong hypercolor
interaction. $\su$ breaking is due to vacuum misalignment$^7$.
\item Top-Mode Standard Models$^8$: in these models it is assumed
that the strength of some strong {\it short-distance} interaction
(perhaps a spontaneously broken gauge theory) is tuned close to the
critical value for chiral symmetry breaking$^9$.
\end{itemize}
If the intrinsic scale of the high-energy interactions (either the
hypercolor interactions in the first case or the short-distance
interactions in the second) is not too much larger than 1 TeV, the
properties of the lightest spin-0 isospin-0 scalar can differ
substantially from the standard model predictions.

\medskip
{\bf If a (iso-scalar) resonance which couples to $WW$ is discovered, how will
we know if it is THE Higgs? }
\medskip

In this talk I will report on the first steps in an analysis of this
question. In particular, I will briefly describe the calculation of
the leading non-analytic corrections to the decay width and to
$W_LW_L$ scattering. The details of the calculations may be found in
ref.  5.

\vfill\eject

\section{The Effective Lagrangian}

We wish to describe a theory in which, in addition to the Goldstone
Bosons (which are, in the sense of the equivalence theorem$^{10}$, the
longitudinal weak gauge bosons) has an iso-singlet scalar field $H$
which is much lighter than any other states in the theory. (Note that,
in this way, the theory is very different than QCD. The ``sigma''
particle in QCD, to the extent it is a distinguishable resonance, is
{\it not} lighter than other resonances.) The lowest-order
interactions of $H$ coupled to the Goldstone Bosons of $SU(2)_L \times
SU(2)_R$ symmetry breaking are summarized by the following effective
low-energy Lagrangian:
\be
\cL = \frac{1}{4} (v^{2} + 2 \xee v H + \xpr H^{2} + \xdpr
\frac{H^{3}}{6 v} )\; \trace\, (\pdmu \yu^{\dagger}\PDmu\yu) \;\;  + \;\;
\cL_{H},  \label{eq:Lagrangian}
\ee
\be
\cL_{H} = \frac{1}{2} (\pdmu H)^2 -  \frac{m^2}{2} H^2 -
    \frac{\ctri v}{3 !}H^3 - \frac{\ctes}{4 !}H^4 , \label{eq:pot}
\ee
where,
\be
\yu = \exp \left(\frac{i \vw \cdot \pauli}{v} \right) \;\;\;\; , \;\;\;
       \trace \,  (\tau^{a} \tau^{b}) = 2 \delta^{a b},
\ee
$v \approx 250$ GeV, and the $\vw$ are the ``eaten'' Goldstone Bosons.

The ordinary linear sigma-model corresponds to the limit:
\be
  \xee \, , \, \xpr \; = \; 1
   \;\;\; , \;\;\;   \xdpr \; = \; 0 \label{eq:smi}
\ee
and
\be
\ctri\, , \, \ctes \; = \; \frac{3 m^2}{v^2} \ \ . \label{eq:smii}
\ee
For a non-standard Higgs, we expect deviations from these
values of order $v^2/\Lambda^2$.

\section{Phenomenology}

We begin by considering the width of the non-standard Higgs Boson.
At tree-level, we find:
\be
\Gamma_{H}^{(0)}=\frac{3m^{3}}{32\pi v^{2}}\xi^{2} \label{eq:treewidth}
\ee
Note that \xee\ is the only parameter which appears$^{11,12}$.

The other parameters in eq.~(\ref{eq:Lagrangian}) appear at
one-loop. As usual, loops induce infinities which can be absorbed in
the effective Lagrangian in the traditional way$^{13}$: namely the
infinities associated with non-derivative interactions are absorbed in
the renormalization of the scalar self couplings in
eq.~(\ref{eq:pot}), while the ones associated with vertices involving
derivatives are absorbed in the counterterms of order $p^{4}$. In
general, these introduce further unknown parameters in our
amplitudes. We compute the leading corrections in the $\overline{MS}$
scheme, setting the $\cO(p^4)$ counterterms to zero when the
renormalization scale $\mu$ is equal to $\Lambda$.  These results
include the so-called ``chiral logarithms'', which are the leading
(non-analytic) contributions if $p^2/\Lambda^2$ is sufficiently
small$^{14}$, and in any case are expected to be comparable to the
full $\cO(p^{4})$\ corrections$^{13}$. In addition, when the
parameters take the values of the linear sigma model
eqs.~(\ref{eq:smi},\ref{eq:smii}), the $\mu$-dependence disappears (as it must
for a renormalizable theory) and our results reduce to
those previously computed in the standard Higgs model.

The one-loop corrections to the Higgs boson decay width in
eq.~(\ref{eq:treewidth}), written as $\Gamma_{H}= \Gamma_{H}^{(0)}+
\Gamma_{H}^{(1)}$, are
\bear
\lefteqn{\frac{\Gamma_{H}^{(1)}}{\Gamma_{H}^{(0)}} = \frac{1}{8\pi^{2}}
  \left\{ \frac{m^{2}}{v^{2}}(1+L)+\frac{\xpr\ctri}{2\xi}\left[
  \frac{\pi}{\sqrt{3}}-1 \right] + \frac{\ctri^{2}v^{2}}{4m^{2}}
   \left[ 1-\frac{2\pi\sqrt{3}}{9}\right]  +\frac{\xdpr
      m^{2}}{\xi v^{2}} L \right. } \nonumber\\ & &
\left. \mbox{}+ \xpr \frac{3m^{2}}{2 v^{2}} \left[\frac{1}{3} +L\right]
      +  \frac{\xi^{2}m^{2}}{2v^{2}} \left[\frac{\pi^{2}}{6}
  -4-5L \right]- \frac{\xi\ctri}{2}\left[\pi\sqrt{3}-3-\frac{2\pi^{2}}{9}
  \right]   \right\} \label{eq:hwidth}
\eear
where $L=1-\ln (m^{2}/\mu^{2})$. While this result is $\mu$-dependent,
we can estimate the effect of higher-order interactions by setting
$\mu=\Lambda$. In the linear sigma model limit our calculation
reproduces the one-loop result of ref. 15:
\be
\frac{\Gamma_{H}^{(1)}}{\Gamma_{H}^{(0)}} = \frac{m^{2}}{2\pi^{2}
v^{2}} \left( \frac{19}{16} -\frac{3\sqrt{3} \pi}{8} + \frac{5
\pi^{2}}{48} \right)
\ee

Next we consider the effects of a non-standard Higgs on Goldstone
Boson scattering.  The tree-level amplitude for \wz\ is
\be
\cA_{tr} = \frac{s}{v^{2}} - \left(\frac{\xi^{2}}{v^{2}}\right)
 \frac{s^{2}}{s-m^{2}- \Sigma(s)}  \label{eq:tree}
\ee
The calculation of the one-loop corrections is straightforward,
though somewhat lengthy. The full analytical expressions
may be found in ref. 5.
At energies small compared to the mass of the Higgs, the one-loop amplitude
is:
\be
\cA(s,t,u) = \frac{s}{v^2} + \frac{1}{(4\pi v^{2})^{2}}\; \cT
 + \xi^2  \frac{s^2}{m^2 v^2}  \label{eq:lowexp}
\ee
\bear
\cT & = & \frac{s^2}{2} \ln\frac{\mu^2 }{-s} + \frac{t}{6} (s +2t)
     \ln \frac{\mu^2}{-t} + \frac{u}{6} (s+ 2u)\ln\frac{\mu^2}{-u}
\nonumber\\*[0.1 cm]  & & \mbox{} +
   s^2 \, P \; + \; Q\, (t^2 +u^2 ) \; + \, R\ln \frac{m^{2}}{\mu^2}
\eear
where
\bear
 P & = & \frac{5}{9} + 2  \xi \xdpr  + \xi^2 \left( \frac{7}{2} \xpr +
     \frac{22}{9} \right)  - \frac{65}{9} \xi^4
   + \frac{\xi \ctri}{2 \lambda^\prime} \left(\xpr - \frac{\xi^2}{2}\right)
  + \xi^2  \frac{\ctri^2}{8 {\lambda^\prime}^{2}} \left(
      \frac{\pi}{\sqrt{3}} - 2  \right) 
\\[0.1 cm]
Q  & = &  \frac{13}{18} -  \frac{11}{9}\xi^{2} + \frac{5}{18} \xi^{4}
\\[0.1 cm]
R  & = & s^{2}\, \left[ \frac{37}{6}\xi^4 - \xi^2 \left( \frac{10}{3}+ 2
           \xpr \right) +2 \xi \xdpr - \frac{\xpr^2}{2}  \right]
  + \frac{\xi^2}{3} \left( 2 - \xi^2 \right) \left(t^2 +u^2 \right)
\eear
where $\lambda^\prime= m^2/2 v^2$.
Also in eq.~(\ref{eq:lowexp}), since $m < 4 \pi v$, we have retained the
$1/m^2$ correction to this order in the momentum expansion.  In the
linear sigma model limit and to leading order in $s/m^2$ this amplitude
explicitly agrees with that of ref. 16.

The differential cross section for longitudinal gauge Boson
scattering is obtained from the amplitude by
\be
\frac{d\sigma}{dt}=\frac{1}{16 \pi s^{2}}\left| \cA \right|^{2}
\ee
where $\cA =  \cA_{tree} +\cA_{loop}$.
Since we neglected higher order corrections, we have
\be
| \cA |^{2} = |\cA_{tree}|^{2} + 2\left\{\, \mbox{\rm
Re}\,( \cA_{tree})\mbox{\rm Re}\,(\cA_{loop})+ \mbox{\rm
Im}\,(\cA_{tree})\mbox{\rm Im}\,(\cA_{loop})\, \right\}\ \ . \label{eq:68}
\ee
The total cross section is then
\be
\sigma_{tot}\/(s)=\int_{-s}^{0}dt\,\frac{d\sigma}{dt}(s,t)
\ee

The amplitude for $W_L^+ W_L^-$ scattering can be
calculated from those given above:
\be
 \cA (W_L^{+}W_L^{-} \rightarrow W_L^{+}W_L^{-}) =  \cA(s,t,u) +
\cA(t,s,u)\; \; .
\ee
\begin{figure}[htb]
\vspace{-0.2in}
\epsfysize 8cm
 \centerline{\epsffile{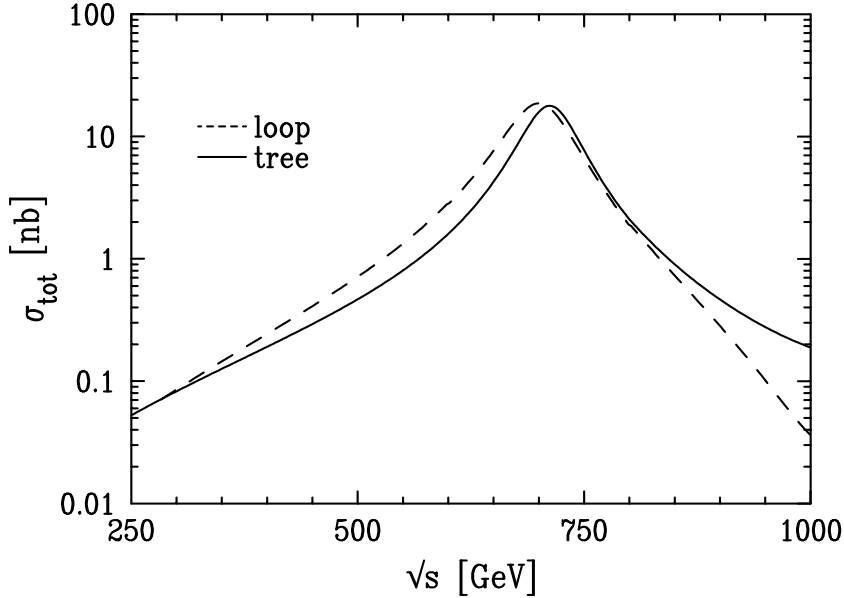}}
\caption{The total cross section for $W^{+}_{L} W^{-}_{L}
\rightarrow W^{+}_{L} W^{-}_{L}$, for a non-standard Higgs with mass
$m = 718$ GeV, and using the values in eq.~(19) as a function
of  $s$. Solid lines correspond to tree level and dashed lines  to
one-loop. }
 \label{fig:fig7a}
  \end{figure}

In Fig.~1 we show the total cross section for the $W_L^+ W_L^-
\rightarrow W_L^+ W_L^-$\ channel as a function of $s$
for a Higgs mass of $m = 718$ GeV, with the parameters
\be
\xi=0.62\;\; ,\; \xpr=-0.21\;\; ,\;\xdpr=0.71\;\; ,\;
\ctri= 18.26\;\; ,\; \ctes= 4.79\ \ ,      \label{eq:vl}
\ee
and $\Lambda=2.2$ TeV.  These parameters are motivated by a composite
Higgs model based on an $SU(4)/Sp(4)$ symmetry structure$^5$. The
solid line corresponds to tree level and the dashed lines to one-loop.
The corresponding curves for a Standard model Higgs with the same mass
are shown in Fig.~2.  The sharp fall in the cross section in the
region above the peak in Fig.~1 can be understood by noticing that for
$\xi < 1$\ the tree amplitude in eq.~(\ref{eq:tree}) vanishes at some
energy greater than $m^2$ (if one does not include a finite
width). This only signals that higher order effects are expected to be
significant there. Also, far above the peak the amplitude presented is
not trustworthy due to the breakdown of the expansion in powers of
$1/\Lambda$.
\begin{figure}[htb]
\vspace{-0.2in}
 \epsfysize=8cm  \centerline{\epsfbox{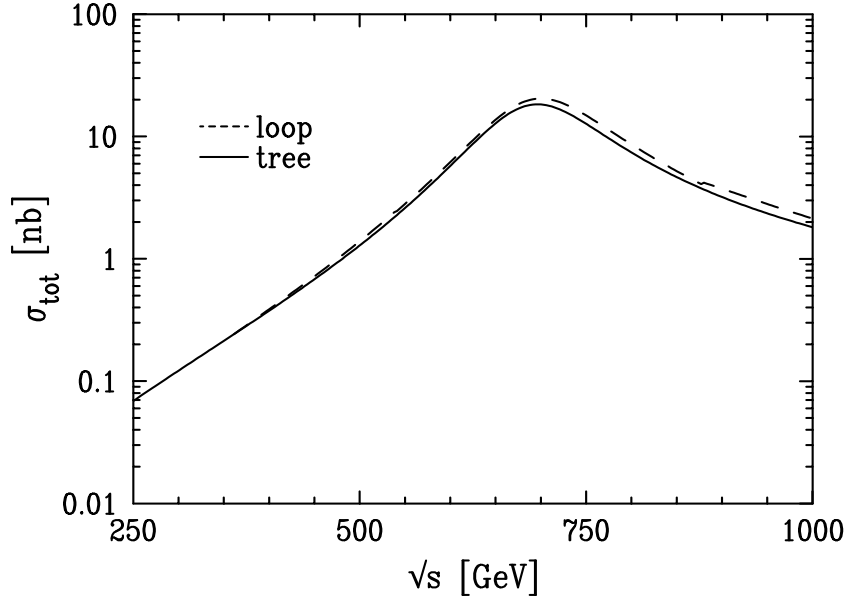}}
  \caption{The total cross section for $W^{+}_{L} W^{-}_{L}
\rightarrow W^{+}_{L} W^{-}_{L}$ in the Standard Model with a Higgs mass
$m = 718$ GeV, as a function of  $s$. Solid lines correspond to tree
level and dashed lines to one-loop.}
  \label{fig:fig7b}
  \end{figure}

Qualitatively, however, for gauge-Boson scattering below a TeV, the
width and shape of the peak appear to be the most important features
differentiating a standard from a non-standard Higgs resonance.

The cross sections discussed above are not directly measurable in
hadron colliders like the LHC; one must first convolute them with the
\ww\ luminosities inside the proton. A more detailed study of how well the
LHC be able to differentiate a standard from a non-standard Higgs can
only be answered after detailed analysis of a specific detector. This
question is currently under investigation.

\section{Acknowledgements}

R.S.C. thanks the organizers for a stimulating workshop.  R.S.C.
acknowledges the support of an Alfred P. Sloan Foundation Fellowship,
an NSF Presidential Young Investigator Award, a DOE Outstanding Junior
Investigator Award, and a Superconducting Super Collider National
Fellowship from the Texas National Research Laboratory Commission.
This work was supported in part under NSF contract PHY-9057173 and DOE
contract DE-FG02-91ER40676, and by funds from the Texas National
Research Laboratory Commission under grant RGFY92B6.

\end{document}